\documentclass[pra,graphicx,twocolumn,groupedaddress,superscriptaddress]{revtex4-2}%

\usepackage{CJK}
\usepackage{amssymb}
\usepackage{amsmath}
\usepackage{dcolumn}
\usepackage{graphicx}
\usepackage{float}
\usepackage{epstopdf}
\usepackage{mathrsfs}
\usepackage{appendix}
\usepackage{booktabs}
\usepackage{color}
\usepackage[table]{xcolor} 
\usepackage{url}
\usepackage[colorlinks]{hyperref}
\hypersetup{%
	plainpages=true,
	breaklinks=true,
	hypertexnames=false,
	pageanchor=true,
	colorlinks=true,
	linkcolor={blue},
	citecolor={blue},
	urlcolor={blue},
	anchorcolor={black}
}

\makeatletter
\renewcommand{\maketag@@@}[1]{\hbox{\m@th\normalsize\normalfont#1}}%
\makeatother
\allowdisplaybreaks[4]

\begin{document}
\preprint{APS/123-QED}
\title{Parallel distributed quantum gates for dual-species quantum emitters}
\author{Zhihao Xie}
\affiliation{MIIT Key Laboratory of Semiconductor Microstructure and Quantum Sensing, School of Physics,  Nanjing University of Science and Technology, Nanjing {\rm 210094}, China}
\author{Adam Miranowicz}
\affiliation{Institute of Spintronics and Quantum Information, Faculty of Physics, Adam Mickiewicz University, PL-61-614 Pozna\'n, Poland}
\affiliation{Center for Quantum Computing, RIKEN, Wakoshi, Saitama 351-0198, Japan}
\author{Zhenhua Li}
\affiliation{MIIT Key Laboratory of Semiconductor Microstructure and Quantum Sensing, School of Physics,  Nanjing University of Science and Technology, Nanjing {\rm 210094}, China}
\affiliation{Engineering Research Center of Semiconductor Device Optoelectronic Hybrid Integration in Jiangsu Province, Nanjing {\rm210094}, China}
\author{Tao Li}
\email[]{tao.li@njust.edu.cn}
\affiliation{MIIT Key Laboratory of Semiconductor Microstructure and Quantum Sensing, School of Physics,  Nanjing University of Science and Technology, Nanjing {\rm 210094}, China}
\affiliation{Engineering Research Center of Semiconductor Device Optoelectronic Hybrid Integration in Jiangsu Province, Nanjing {\rm210094}, China}
\author{Franco Nori}
\affiliation{Center for Quantum Computing, RIKEN, Wakoshi, Saitama 351-0198, Japan}
\affiliation{Physics Department, The University of Michigan, Ann Arbor, Michigan 48109-1040, USA}

\date{\today}

\begin{abstract}
We propose a parallel protocol for implementing distributed nonlocal quantum gates between spatially separated stationary qubits encoded in dual-species quantum emitters~(i.e., color-center and superconducting qubits).~By utilizing entangled photon pairs with distinct frequencies as a quantum data bus, our approach connects spatially separated devices without requiring quantum frequency conversion or preshared entanglement, while maintaining an always-ready and resource-efficient property for distributed quantum computing and networks.  Furthermore, we demonstrate the feasibility of implementing \emph{parallel distributed nonlocal} quantum gates on multiple pairs of spatially separated qubits using a single high-dimensional entangled photon pair, which directly benefits from the enhanced quantum capacity provided by optical qudit encoding. Our protocol establishes a scalable and practically implementable framework for distributed quantum networks, potentially enabling the development of future large-scale quantum computing architectures.
\end{abstract}

\maketitle

\section{Introduction} \label{sec:1}
Quantum computing harnesses parallel quantum power~\cite{ladd2010quantum} and has the potential to solve problems beyond the capacity of classical computers~\cite{Georgescu2014Quantumsimulation,Long2001Grover,zhou2020limits}. Quantum advantages have been demonstrated for specific applications with noisy-intermediate-scale quantum computing~\cite{Arute2019Quantum,Zhong2020,Madsen2022,Wei2022quantum}. However, it is challenging to directly scale up these physical platforms to construct a large-scale quantum computer due to inevitable errors and noise~(i.e., crosstalk) within a finite device~\cite{Cheng2023Noisy}. Distributed quantum computing and networks provide an efficient approach to tackle this problem by modulating spatially separated devices~\cite{Cirac99Distributed,Nemoto2014Photonic,reiserer2015cavity}. 
To form a scalable quantum computer cloud~\cite{Sheng2018Blind}, efficient and faithful \emph{nonlocal} quantum gates operating on two qubits situated in different devices are necessary to connect them together for implementing universal quantum information processing~\cite{Lim2005Repeat,Jiang2007Distributed,zheng2010arbitrary}.

Distributed quantum computing and networks inherently decouple inter-device interactions and thus prevent the crosstalk between qubits of different devices~\cite{wehner2018quantum}, making nontrivial quantum gates implemented by direct interactions impossible for spatially separated qubits~\cite{Cirac99Distributed,Nemoto2014Photonic,reiserer2015cavity,Sheng2018Blind, Lim2005Repeat,Jiang2007Distributed,zheng2010arbitrary}. Single photons, acting as a data bus, interact sequentially with two stationary qubits and then produce indirect interactions between these qubits, facilitating the desired nontrivial quantum gates~\cite{Duan2005Robust,Lin2006One-step,hu2008giant, Cohen2018, Daiss2021quantum-logic}. These protocols function for indistinguishable atomic qubits and solid-state qubits with {a} finite and rectifiable discrepancy~\cite{Vittorini2014Entanglement,Stockill2017Phase-Tuned, Dibos2018Atomic,Zhai2022Quantum}, and show the distinct property of being always ready for a gate operation on spatially separated qubits without shared entanglement~\cite{Daiss2021quantum-logic}. Furthermore, postselection is exploited to increase their fidelities at the cost of success probabilities. In principle, such protocols can be generalized to implement quantum gates on dual-species qubits, once  {a} complicated quantum frequency conversion of photons is achieved~\cite{Leent2020Long-Distance,Zhou2024Long-Lived,Knaut2024Entanglement}. 

Quantum-gate teleportation transfers nontrivial quantum gates on two local qubits within the same device to two spatially separated qubits across different devices~\cite{Gottesman1999Demonstrating}, using pre-shared entanglement connecting two devices, classical communication, and local operations. The distance between two devices can be significantly larger when compared to protocols using single photons, since the auxiliary  pre-shared entanglement can be generated by connecting short-distance entanglement with quantum repeaters~\cite{Sangouard2010RMP,sheng2013hybrid,Wang2012QR,Munro2015Inside}. Recently, several proof-of-principle experiments of this approach have demonstrated nonlocal quantum gates on spatially separated qubits~\cite{Chou2018Deterministic,Wan2019Quantum,qiu2023deterministic,Liu2024Nonlocal,inc2024distributed,Feng2025Chip-to-Chip,main2024distributed}. While many important works primarily focus on heralded entanglement generation or state transfer~\cite{liu2021heralded,barz2010heralded,usmani2012heralded,hofmann2012heralded,bernien2013heralded,riedinger2018remote,wengerowsky2018entanglement,yang2013entanglement}, a universal \emph{nonlocal} gate, such as a \emph{distributed} controlled-NOT~(CNOT) gate, provides a universal primitive that can be directly embedded into algorithm-level quantum circuits involving spatially separated qubits~\cite{main2024distributed}. The number of pre-shared entangled pairs imposes an upper bound for the performance of nonlocal quantum gates and reduces the accessible state space within each device~\cite{Chou2018Deterministic,Wan2019Quantum,qiu2023deterministic,Liu2024Nonlocal,inc2024distributed,Feng2025Chip-to-Chip,main2024distributed}.

Here, we propose a parallel protocol to implement \emph{distributed nonlocal quantum gates} for stationary qubits encoded in dual-species quantum emitters. This work focuses on \emph{distributed quantum operations} that, together with local gates, enable universal quantum computation. An entangled photon pair source generates two entangled photons of different frequencies~\cite{Yan2011PRL,Dong2017OE,Lu2019NatPhys,Zhao2014PRL,Yang2022PRB,APL2021QDReview}. The entangled photons, rather than single photons, serve as a quantum data bus connecting two spatially separated devices when their frequencies match the stationary qubits, enabling deterministic interfaces between single photons and stationary qubits~\cite{Beukers2024Remote-Entanglement}. The protocol is \emph{always ready for distributed nonlocal quantum gates} on two spatially separated dual-species qubits, \emph{requiring neither quantum frequency conversion}~\cite{Leent2020Long-Distance,Zhou2024Long-Lived,Knaut2024Entanglement} \emph{nor pre-shared entanglement between devices}~\cite{Gottesman1999Demonstrating,Sangouard2010RMP,sheng2013hybrid,Wang2012QR,Munro2015Inside}. In particular, the protocol is formulated to realize \emph{nonlocal} CNOT gates between spatially separated qubits and to enable their parallel execution, which is directly relevant for distributed quantum algorithms. Furthermore, we demonstrate parallel distributed quantum gates operating simultaneously on multiple qubit pairs of dual-species qubits across different devices. This can be achieved by distributing a single high-dimensional entangled photon pair~\cite{Erhard2020high-dimensional,deng2017quantum,Liu2020Low-Cost,Du2025Generation},  benefiting directly from high-capacity quantum channels enabled by optical qudit encoding. These results position our protocol as a useful building block for distributed quantum computing and quantum network architectures.

\begin{figure}[t!]
	\includegraphics[width=8.6 cm]{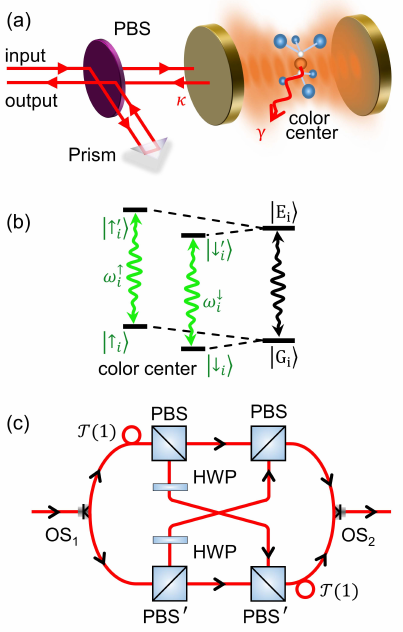}
	\caption{(a) Schematic of the CPF gate based on a color center~(i.e., SiV$^-$ or GeV$^-$) coupled to a single-sided cavity. Here, PBS stands for a polarizing beam splitter, and $\gamma$ ($\kappa$) is the color center (cavity) decay rate. (b) Relative-level structure and optical transitions of {the} negatively charged
SiV$^-$ and GeV$^-$ color centers. The dipole transitions coupling {the ground branch~($|G_i\rangle$) states $|\!\uparrow_{i}\rangle$ and $|\!\downarrow_{i}\rangle$ to  the  excited branch~($|E_i\rangle$)  states $|\!\uparrow_{i}'\rangle$ and $|\!\downarrow'_{i}\rangle$) for $i=A,~B$ are $H$ polarized with frequencies $\omega_i^\uparrow$ and $\omega_i^\downarrow$ of the SiV$^-$ and GeV$^-$ color centers, respectively. (c) Schematic of the gate U$_s$.}}\label{fig1}
\end{figure}

\section{The protocol and its physical implementation}
The parallel distributed protocol begins with an entangled photon pair source that generates two photons of different frequencies. These photons act as a quantum data bus, connecting two spatially separated devices, where their frequencies match the dual-species qubits. This enables efficient interactions between single photons and qubits, facilitating hybrid CNOT gates.

The distributed CNOT gate~(i.e., with the control qubit $s_1$ in one species and the target qubit $s_2$ in the other species) can be implemented using the entangled photon pair \emph{A} and \emph{B} as an intermediate resource: First, a hybrid CNOT gate is performed with one photon acting as the control and $s_2$ as the target. Next, the other photon from the entangled pair entangles with $s_1$. Finally, single-photon measurements in a conjugate basis herald the successful completion of the distributed CNOT gate on the spatially separated dual-species qubits, up to single-qubit operations on $s_1$ and $s_2$.

Moreover, by utilizing a single high-dimensional entangled photon pair, the parallel implementation of distributed CNOT gates on multiple pairs of dual-species qubits can be achieved. After completing the distributed CNOT gate on the first pair of dual-species qubits, the photon-pair state is rearranged to
effectively reset the entangled photons. This enables the
subsequent CNOT gate to be implemented on the next qubit pair
through a similar process, using the same photon pair.

For color centers in diamond~\cite{Awschalom2018Quantum}, a controlled-polarization flip~(CPF) gate enables an effective interface between single photons and stationary qubits, which can be implemented by coupling a four-level negatively charged silicon vacancy~(SiV$^-$) or germanium  vacancy~(GeV$^-$) color center to a single-sided nanocavity~\cite{Nguyen2019Network,Zifkin2024Lifetime} and properly tuning the photon polarization with a prism and a polarizing beam splitter~(PBS), shown in Fig.~\ref{fig1}. The PBS transmits $H$-polarized photons and reflects $V$-polarized photons. An $H$-polarized photon nearly resonant with the transition $|\!\!\!\uparrow_{i}\rangle\leftrightarrow|\!\!\!\uparrow_{i}'\rangle$ of frequency $\omega_i^\uparrow$ acquires  {no} phase shift, when it is scattered by the cavity that interacts with the SiV$^-$ or Ge$^-$ {center} in {the} state $|\!\!\!\uparrow_{i}\rangle$ in the {strong-coupling} regime. In contrast, the photon acquires a {phase shift} ${\pi}$ when the SiV$^-$ or GeV$^-$ {center} is in {the} state $|\!\!\downarrow_{i}\rangle$, which couples to {the} state $|\!\!\downarrow_{i}'\rangle$ with transition frequency $\omega_i^\downarrow=\omega_i^\uparrow+\Delta_i$ and significantly detunes from the cavity mode by $\Delta_i$. For a photon in a general pure state $|\phi_p\rangle=\alpha|D\rangle+\beta|A\rangle$~[e.g., $|D\rangle=(|V\rangle+|H\rangle)/\sqrt{2}$, $|A\rangle=(|V\rangle-|H\rangle)/\sqrt{2}$] and a color center in the state $|\phi_s\rangle=\alpha'|\!\!\uparrow_{i}\rangle+\beta'|\!\!\downarrow_{i}\rangle$, the output state of the photon and the color center after the interaction within the CPF gate evolves to~\cite{Zhou2023parallel}  
\begin{eqnarray}
	|\Phi\rangle=\alpha'|\!\!\uparrow_{i}\rangle|\phi_p\rangle+\beta'\hat{X}_p|\!\!\downarrow_{i}\rangle|\phi_p\rangle,  \label{eq-cpf}
\end{eqnarray} 
where the operator $\hat{X}_p=|D\rangle\langle A|+|A\rangle\langle D|$ completes the polarization-flip operation $|D\rangle\leftrightarrow|A\rangle$ when the color center is in the state $|\!\!\downarrow_{i}\rangle$.

\begin{figure*}[t!]
	\includegraphics[width=17.2 cm]{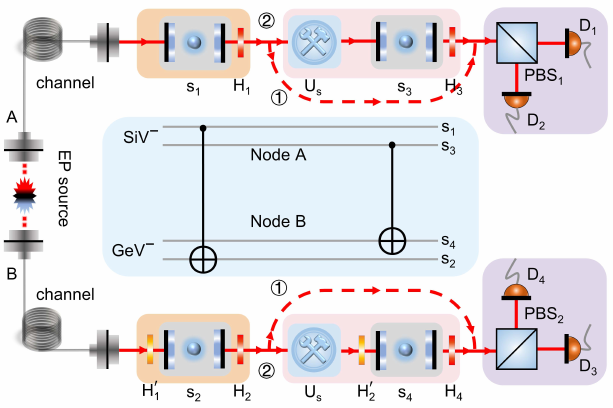}
	\caption{Schematics of distributed CNOT gates on spatially separated dual-species qubits. The SiV$^-$ ($s_1$, $s_3$) centers  are the control qubits and the GeV$^-$~($s_2$, $s_4$) centers  are the target qubits, encoded in two ground states~($\vert\!\!\uparrow_{A,B}\rangle$, $\vert\!\!\downarrow_{A,B}\rangle$). H$_i$~($i=1, \ldots, 4$) performs the Hadamard transformation, while H$'_{1,2}$ performs a Hadamard-like transformation. PBSs are polarizing beam splitters. $\rm D_u$ (u=1, \ldots, 4) are detectors, $U_s$ is a gate that performs $\mathcal{T}(1)|H\rangle\leftrightarrow\mathcal{T}(0)|V\rangle$, while \textcircled{1} and \textcircled{2} indicate different optical paths.}\label{fig2}
\end{figure*}

The schematics for implementing the distributed CNOT  gates on dual-species stationary qubits located in two spatially separated nodes is shown in Fig.~\ref{fig2}. 
The stationary qubits in node \emph{A}~(\emph{B}) are encoded on the SiV$^-$~(GeV$^-$) electron spins, and the corresponding CPF gates are referred to as $s_1$ and $s_3$~($s_2$ and $s_4$) for simplicity of notation  hereafter. An  entangled photon (EP) source prepares a pair of photons $|\varphi_p\rangle=(|H_AH_B\rangle+|V_AV_B\rangle)/\sqrt{2}$. The subscripts $A$ and $B$ denote  photons sent to nodes \emph{A} and \emph{B}, which are of frequencies around $\omega^\uparrow_A$ and $\omega^\uparrow_B$ of the transitions of the SiV$^-$ and GeV$^-$ color center, respectively. 

Suppose now that the stationary qubits $s_1$ and $s_2$ are initialized in the normalized states $\vert \varphi_1\rangle=\alpha_1\left\vert \uparrow_A\right\rangle+\beta_1\left\vert\downarrow_A\right\rangle$ and  $\vert\varphi_2\rangle=\alpha_2\left\vert\uparrow_B\right\rangle+\beta_2\left\vert\downarrow_B\right\rangle$, within the ground branch $|G_i\rangle$~\cite{Nguyen2019Network}. On arriving at the node \emph{B}, photon \emph{B} passes through {H}$'_1$~[i.e., a half-wave plate (HWP) oriented $-\pi/8$] undergoing a Hadamard-like transformation (i.e., $|H\rangle \rightarrow -|A\rangle$ and $|V\rangle \rightarrow |D\rangle$), which evolves $|\varphi_p\rangle$ to $|\varphi'_p\rangle=(|D_AH_B\rangle+|A_AV_B\rangle)/\sqrt{2}$. Then photon \emph{A} is directed to interact with $s_1$ and photon B is directed to interact with $s_2$, before and after which  {the} Hadamard transformation~[i.e., $\left|\uparrow_{B}\right\rangle \rightarrow ( \left|\uparrow_{B}\right\rangle+ \left|\downarrow_{B}\right\rangle)/\sqrt{2}$ and $\left|\downarrow_{B}\right\rangle\rightarrow (\left|\uparrow_{B}\right\rangle- \left|\downarrow_{B}\right\rangle)/\sqrt{2}$] is performed on $s_2$. 
The combined states of photons \emph{A} {and} \emph{B} and qubits $s_1$ {and} $s_2$ evolves into 
\begin{eqnarray}
	|\Phi_{1}\rangle&=&\frac{1}{2}
\big(\alpha_1\hat{X}_{2}|\uparrow_{s1}\rangle|D\rangle|H\rangle
+\beta_1\hat{X}_{2}|\downarrow_{s1}\rangle|A\rangle|H\rangle\nonumber\\
&& 
+\alpha_1\hat{I}_{2}|\uparrow_{s1}\rangle|A\rangle|V\rangle
+\beta_1\hat{I}_{2}|\downarrow_{s1}\rangle|D\rangle|V\rangle\big)|\varphi_2\rangle,
\label{eq1}
\end{eqnarray}
where $\hat{X}_2=|\!\uparrow_{s2}\rangle\langle \downarrow_{s2}\!|+|\!\downarrow_{s2}\rangle\langle \uparrow_{s2}\!|$ is  {the} bit-flip operator for $s_2$ and $\hat{I}_2$ is the identity operator for $s_2$.

To complete the distributed CNOT gate on $s_1$ and $s_2$, photons \emph{A} and \emph{B}   {pass} through H$_1$ and H$_2$, respectively; and then each photon  undergoes the Hadamard transformation~(i.e., $|H\rangle \rightarrow |D\rangle$ and $|V\rangle \rightarrow -|A\rangle$) and is directed to the corresponding measurement units within the basis $\{|H\rangle, |V\rangle\}$ by {the} optical {path} {\footnotesize \textcircled{1}}, which evolves the combined system into 
\begin{eqnarray}
	|\Phi_{\rm 2}\rangle  &=& \frac{1}{2}\Big(\hat{U}_{c_1}|HH\rangle+\hat{U}_{c_2}|HV\rangle-\hat{U}_{c_3}|VH\rangle \nonumber\\
	 	&& -\hat{U}_{c_4}|VV\rangle\Big)\otimes|\varphi_1\rangle|\varphi_2\rangle,\label{eq3}
\end{eqnarray}
where $\hat{U}_{c_1}$ applies  the CNOT gate on $s_1$ and $s_2$ with
$\hat{U}_{c_1}|\varphi_1\rangle|\varphi_2\rangle=\alpha_1\hat{X}_2|\uparrow\rangle|\varphi_2\rangle+\beta_1\hat{I}_2|\downarrow\rangle|\varphi_2\rangle$, $\hat{U}_{c_2}=\hat{Z}_1\hat{U}_{c_1}$, $\hat{U}_{c_3}=\hat{X}_2\hat{U}_{c_1}$, and $\hat{U}_{c_4}=-\hat{Z}_1\hat{X}_2\hat{U}_{c_1}$. Therefore, when two polarized photons are detected simultaneously, the success of the distributed CNOT gate with the SiV$^-$ qubit $s_1$ as the control  and GeV$^-$ qubit $s_2$ as the target is heralded, up to local operations on $s_1$ and $s_2$.

Our distributed CNOT-gate protocol for dual-species stationary qubits~\cite{Awschalom2018Quantum} can be generalized to implement \emph{parallel distributed} CNOT gates on \emph{multipairs} of qubits with $N$ SiV$^-$ spin qubits in the node \emph{A} as the control and $N$ corresponding GeV$^-$ spin qubits in the node \emph{B} as the target. Here we detail the schematic of two parallel distributed CNOT gates with the transmission of a single pair of photons, shown in Fig.~\ref{fig2}. Assume now that the SiV$^-$ qubits $s_i$ for $i=1,3$  are initialized to the normalized states 
$\vert \varphi_i\rangle=\alpha_i\left|\uparrow_A\right\rangle+\beta_i \left\vert\downarrow_A\right\rangle$, GeV$^-$ qubits $s_i$ for $i=2,4$  are initialized to the normalized states 
$\vert \varphi_i\rangle=\alpha_i\left|\uparrow_B\right\rangle+\beta_i \left\vert\downarrow_B\right\rangle$,  and the EP source prepares an entangled photon pair in the state 
$\vert \phi_{j}\rangle=\frac{1}{\sqrt{2}}\sum_{k=0}^1\mathcal{\hat{T}}_1(k)\hat{\mathcal{T}}_2(k)|\varphi_p\rangle$. The operator $\hat{\mathcal{T}}_l(k)$ introduces a time delay of $k\Delta_t$ on the $l_{\rm th}$ photon  with $\Delta_t$ being the minimum between two neighboring time-bin modes~\cite{Zheng2022Entanglement,Bouchard2022Quantum,Borregaard2020One-Way}.

Photons \emph{A} and \emph{B} pass through the same linear optical elements upon arriving at the nodes \emph{A} and \emph{B},  and then interact with $s_1$ and $s_2$, followed by {the} Hadamard transformations~(H$_1$ and H$_2$), respectively. The combined state of photons \emph{A} {and} \emph{B} and {qubits} $s_1-s_4$ evolves into 
\begin{eqnarray}
	|\Psi_{1}\rangle  &=& \frac{1}{2}{\sum_{k=0}^1}\mathcal{\hat{T}}_1(k)\hat{\mathcal{T}}_2(k)\big(\hat{U}_{c_1}|HH\rangle+\hat{U}_{c_2}|HV\rangle \nonumber\\
	 	&& -\hat{U}_{c_3}|VH\rangle -\hat{U}_{c_4}|VV\rangle\big)|\varphi_s\rangle,\label{eq4}
\end{eqnarray}
which is a direct extension of Eq.~\eqref{eq3} with $|\varphi_s\rangle=|\varphi_1\rangle|\varphi_2\rangle|\varphi_3\rangle|\varphi_4\rangle$. To complete the parallel CNOT gate on {qubits} $s_3s_4$ with the state of {qubits} $s_1s_2$ unchanged, a state exchanging operation, $\mathcal{T}_{i}(1)|H\rangle\leftrightarrow\mathcal{T}_{i}(0)|V\rangle$, is applied on the $i_{\rm th}$ photon $ (i=A,B)$ by passing it through the gate U$_s$ in Fig.~\ref{fig2}. The state  $|\Phi_{\rm out}\rangle$ is transformed into
\begin{eqnarray}
|\Psi'_{2}\rangle  &=& \frac{1}{2}\big(\mathcal{\hat{T}}_1(0)\hat{\mathcal{T}}_2(0)\hat{U}_{c_1}+\mathcal{\hat{T}}_1(0)\hat{\mathcal{T}}_2(1)\hat{U}_{c_2} \nonumber\\
&&-\mathcal{\hat{T}}_1(1)\hat{\mathcal{T}}_2(0)\hat{U}_{c_3} -\mathcal{\hat{T}}_1(1)\hat{\mathcal{T}}_2(1)\hat{U}_{c_4}\big)|\varphi_p\rangle|\varphi_s\rangle,\;\;\;\label{eq5}
\end{eqnarray}
where photons $A$ and $B$ are reset into their initial entangled state $|\varphi_p\rangle$ with different time-bin modes  {to} distinguish the controlled operations on {qubits} $s_1$ {and} $s_2$ and can be used to complete the CNOT gate on {qubits} $s_3$ {and} $s_4$ in optical paths {\footnotesize \textcircled{2}} with the same configuration as that used for completing the CNOT gate on qubits $s_1$ {and} $s_2$, shown in Fig.~\ref{fig2}. 

The combined state of the system before photons \emph{A} {and} \emph{B} entering the measuring units evolves into 
\begin{eqnarray}
|\Psi_{\rm 2}\rangle &=& \big(\mathcal{\hat{T}}_1(0)\hat{\mathcal{T}}_2(0)\hat{U}_{c_1}+\mathcal{\hat{T}}_1(0)\hat{\mathcal{T}}_2(1)\hat{U}_{c_2} \nonumber\\
    &&-\mathcal{\hat{T}}_1(1)\hat{\mathcal{T}}_2(0)\hat{U}_{c_3} -\mathcal{\hat{T}}_1(1)\hat{\mathcal{T}}_2(1)\hat{U}_{c_4}\big)
        \big(\hat{U}'_{c_1}|HH\rangle\nonumber\\
    && +\hat{U}'_{c_2}|HV\rangle -\hat{U}'_{c_3}|VH\rangle -\hat{U}'_{c_4}|VV\rangle\big)|\varphi_s\rangle,
	\label{eq6}
\end{eqnarray}
where $\hat{U}_{c_1}'$ applies  {the} CNOT gate on $s_3$ and $s_4$, $\hat{U}'_{c_2}=\hat{Z}_3\hat{U}'_{c_1}$, $\hat{U}'_{c_3}=\hat{X}_4\hat{U}'_{c_1}$, and $\hat{U}'_{c_4}=-\hat{Z}_3\hat{X}_4\hat{U}'_{c_1}$. Therefore, the parallel distributed CNOT gates on $s_1s_2$ and $s_3s_4$, with two GeV$^-$ qubits $s_2$ and $s_4$ as the target and two SiV$^-$ qubits $s_1$ and $s_3$ as the control, can be completed, up to local operations on them, when photons \emph{A} and \emph{B} are detected by the time-resolved single-photon detectors D$_i$ $(i=1,...,~4)$.

\begin{figure}[t]
	\includegraphics[width=0.458 \textwidth]{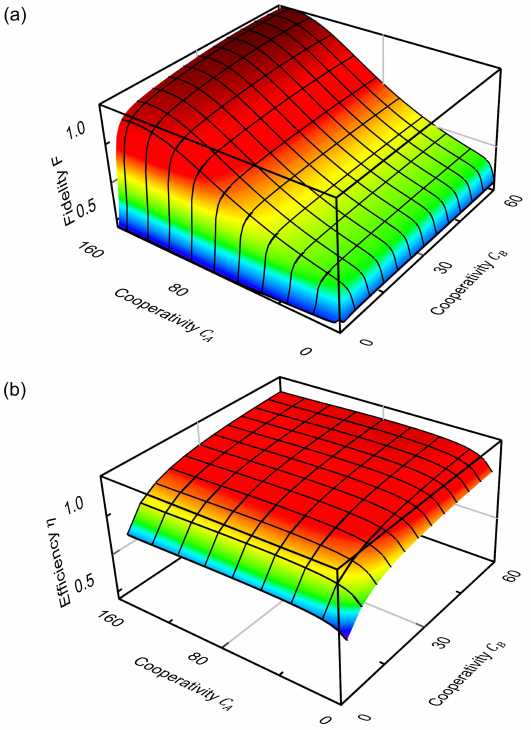}
	\caption{Average fidelity $F$ and efficiency $\eta$ of the parallel distributed CNOT gate vs the color center--cavity cooperativities $C_A$ and $C_B$. }\label{fig3}
\end{figure}

\section{Performance analysis} 
So far, we have described the parallel distributed CNOT gates on dual-species color centers  {for} ideal CPF {gates}, where {the} reflection coefficients $r^\uparrow=1$ and $r^\downarrow=-1$ have been used. However, practical reflection coefficients always deviate from these ideal values, as shown in Appendixes~\ref{App.A} and \ref{App.B}, due to finite detunings and {the color-center and cavity coopreativities $C_j=4g_j^2/\kappa\gamma$}, and can be described as~\cite{scully1997quantum,reiserer2015cavity}
\begin{eqnarray}    
	r^s_j(\omega_j)	=1-\frac{2(i\Delta^s_j+1)}{(i\Delta^s_j +1)(i\Delta^c_j+1)+C_j},
	\label{rcoe}
\end{eqnarray}
where $\Delta^{s}_j=2(\omega^{s}_j-\omega_j)/\gamma$ with the superscript $s=~\uparrow,~\downarrow$
and  $\Delta^c_j=2(\omega^{c}_j-\omega_j)/\kappa$ are {the} effective detunings of the dipole transition of the color center and the cavity mode from the input field frequency~$\omega_j$ in the nodes $j=A$ and $B$, respectively, {where $\gamma$ ($\kappa$) is the color center (cavity) decay rate}.

The average fidelity and efficiency of the parallel distributed CNOT gates with two SiV$^-$ qubits as the controls and {the} two GeV$^-$ qubits as the targets are
\begin{eqnarray}
F&=&\frac{1}{16}\sum_{{m,n}}\int_{0}^{1}\!\!\!\int_{0}^{1}\!\!\!\int_{0}^{1}\!\!\!\int_{0}^{1}|\langle\Psi''_{mn}|\Psi_{mn}\rangle|^2d\alpha_1d\alpha_2d\alpha_3d\alpha_4,\nonumber\\
    \eta&=& \sum_{m,n}\int_{0}^{1}\!\!\!\int_{0}^{1}\!\!\!\int_{0}^{1}\!\!\!\int_{0}^{1}\eta_{c_m} \eta'_{c_n}d\alpha_1d\alpha_2d\alpha_3d\alpha_4,
	\label{eq11}
\end{eqnarray}
where $|\Psi''_{mn}\rangle= \hat{U}_{c_m}\hat{U}'_{c_n}|\varphi_s\rangle$ is the ideal output state of the parallel CNOT gates, while $|\Psi_{mn}\rangle$ is the realistic output state when photons \emph{A} and \emph{B} have been measured to be  $\mathcal{\hat{T}}_A(i)\hat{\mathcal{T}}_B(j)\hat{X_A^k}\hat{X_B^l}|HH\rangle$ for $i,j,k,l\in\{0,1\}$, and the corresponding probability is $\eta_{c_{m}}\eta'_{c_{n}}$ for $[m]_{10}=[ij]_2$ and $[n]_{10}=[kl]_2$. The system evolution and the realistic output state $|\Psi_{mn}\rangle$ have been described in Appendix~\ref{App.A}.

The average fidelity and efficiency as a function of the cooperativities  $C_A$ and $C_B$ 
are shown in Fig.~\ref{fig3}. We  {assume experimentally} accessible parameters~\cite{levonian2022optical}, and assume that photons \emph{A} and \emph{B} are in   {resonance} with one dipole transition of {}{the} SiV$^-$~($s_{1}s_3$) and GeV$^-$~($s_2s_4$) color centers with $\Delta^{\uparrow}_A=\Delta^{\uparrow}_B$=0, respectively, while they are largely detuned from the other dipole transition with $\Delta^{\downarrow}_A=\Delta^{\downarrow}_B=100$~\cite{Zhou2023parallel}. Meanwhile, the normalized cavity detunings are 
($\Delta^{c}_A$, $\Delta^c_B)=(1.5, 0.5)$~\cite{levonian2022optical}. The average fidelity and efficiency can  {reach} $F\simeq0.999$ and $\eta\simeq0.890$ for $C_A=150$ and $C_B=50$, respectively. In addition, the electron-spin coherence time of the SiV$^-$ and GeV$^-$ color centers can be larger than 10 ms~\cite{Sukachev2017Silicon-Vacancy,2024Germanium}, which leads to a fidelity   {decrease} less than $10^{-4}$, since the operation time of our protocol is less than $1$ $\mu$s.

In practice, entangled photon pairs with distinct frequencies can be generated using a variety of well-established optical and atomic platforms~\cite{Yan2011PRL,Dong2017OE,Lu2019NatPhys,Zhao2014PRL,Yang2022PRB,APL2021QDReview}. 
Nondegenerate spontaneous parametric down-conversion and four-wave mixing enable the direct generation of photon pairs at different wavelengths while preserving entanglement in polarization, time-bin, or energy-time degrees of freedom~\cite{Yan2011PRL,Dong2017OE}. 
Such frequency-nondegenerate entanglement has been demonstrated in bulk and atomic-ensemble systems~\cite{Yan2011PRL,Dong2017OE}, as well as in integrated nonlinear photonic platforms, including chip-integrated visible-telecom sources with high brightness and purity~\cite{Lu2019NatPhys}. 
In addition, entanglement between spectrally distinct photons can be generated or processed using time-resolved measurements with active feedforward~\cite{Zhao2014PRL}, frequency-domain Bell-state analysis~\cite{Lingaraju2022Optica}, or auxiliary-photon heralding schemes~\cite{barz2010heralded,Wagenknecht2010NatPhoton}.
Furthermore, deterministic or near-deterministic entangled-photon sources based on semiconductor quantum dots provide on-demand generation of distinguishable entangled-photon pairs with high fidelity~\cite{Yang2022PRB,APL2021QDReview}.

When nonmaximally entangled photon pairs are used, the average fidelity of our protocol decreases as the initial entanglement fidelity $F_0$ of the photon pairs decreases (see Appendix~\ref{App.C}). To enhance the entanglement of the photon pairs, entanglement purification can be employed~\cite{Yan2023Advances}. With a single round of purification, the 	average fidelity can be increased to approximately $F\approx
	0.999$ for an initial fidelity $F_0\approx0.969$. 

\section{Discussion and summary} 
We have described parallel distributed CNOT gates for dual-species quantum emitters and exemplified {its} implementation with GeV$^-$ and SiV$^-$ color centers. In practice, our protocol can be extended to implement parallel CNOT gates for other natural or artificial stationary qubits, shown in Appendix~\ref{App.D}, that interact effectively with single photons encoded in  polarization,  time-bin modes or other degrees of freedom~\cite{Zheng2022Entanglement,Uppu2021quantum,Kono2018Quantum}. It can also be extended to implement distributed CNOT gates on dual-species qubits, beyond single atoms, and even to implement distributed multiqubit gates on multi-species quantum emitters~\cite{Hartung2024quantum-network,Qin2018Exponentially,qin2024quantum,Ramette2022Any-To-Any,gonzalez2024light}. Furthermore, when high-dimensional entangled photon pairs are used~\cite{munro2010quantum,Piparo2020Resource,Lv2024Demonstration,Chen2024High-dimensional}, we can expect {that} the parallel implementation of more pairs of distributed CNOT gates can be directly applied to entangle spatially separated error-correcting logical qubits~\cite{Devitt2013Quantum,Li2024Heralded,han2024protecting}. {}{Note that the parallel-gate capability enabled by high-dimensional photonic encoding could be regarded as a resource-consolidation feature~\cite{munro2010quantum,Piparo2020Resource,Lv2024Demonstration,Chen2024High-dimensional}, rather than as a route toward large-scale simultaneous gate execution. In practical implementations, the achievable degree of parallelism is expected to be constrained by experimental complexity and decoherence, and thus is likely to remain modest within realistic experimental regimes.}

In summary, we have presented a protocol for implementing distributed CNOT gates on spatially separated dual-species quantum emitters. We use an entangled photon pair as a {data bus} to directly connect dual-species quantum emitters without requiring  quantum frequency conversion. Furthermore, we demonstrate that parallel distributed CNOT gates on multiple pairs of stationary qubits can be achieved by transmitting a single photon pair with additional time-bin encoding. {}{Our protocol provides a concrete framework for exploiting hybrid quantum systems in distributed quantum information processing and offers a feasible route toward the development of hybrid quantum technologies.}

\begin{acknowledgments}
This  work  was  supported  by the National Natural Science Foundation of China (Grants No.~11904171 and No.~62221004).
A.M. was supported, within the QEC4QEA project, by the European
High Performance Computing Joint Undertaking (EuroHPC JU) under
Grant No. 101194322 and by the Polish National Centre for Research
and Development (NCBR) under Grant No. DWM/EuroHPC/2023/429/2025. 
F.N. is supported in part by the Japan Science and Technology Agency (JST) [via the CREST Quantum Frontiers program (Grant No. JPMJCR24I2), the Quantum Leap Flagship Program (Q-LEAP), the Moonshot R\&D (Grant No. JPMJMS256E), and the ASPIRE program (Grant No. JPMJAP2513)].
\end{acknowledgments}

\appendix
\section{Practical state evolution of the combined system for implementing the parallel distributed CNOT gates}~\label{App.A}

In the main text, we described the state evolution of the combined system consisting of  photons and stationary qubits using ideal CPF gates. Here, we provide details on the state evolution for realistic CPF gates that incorporate practical single photon-qubit interfaces. An $H$-polarized photon, after scattering by a cavity coupled to a qubit, is changed by a reflection coefficient dependent on the qubit state, leading to the following evolution:
\begin{align}
	|H\rangle|\uparrow_{}\rangle\rightarrow r_{}^\uparrow|H\rangle|\uparrow_{}\rangle, \qquad	|H\rangle|\downarrow_{}\rangle\rightarrow r_{}^\downarrow|H\rangle|\downarrow_{}\rangle.
\end{align} 
Here the state-dependent reflection coefficients are
\begin{align}    
	r^s_j(\omega_j)	=1-\frac{2(i\Delta^s_j+1)}{(i\Delta^s_j +1)(i\Delta^c_j+1)+C_j}, 
	\label{rcoes}
\end{align}
where $\Delta^{s}_j=2(\omega^{s}_j-\omega_j)/\gamma$ with the superscript $s=\uparrow, \downarrow$. Also, $\Delta^c_j=2(\omega^{c}_j-\omega_j)/\kappa$ are effective detunings of the dipole transition of the color center and the cavity mode from the input field frequency~$\omega_j$ in nodes $j=A$ and $B$, respectively, and the $C_j$'s are the color center-cavity cooperativities.

For all stationary qubits initialized in the states $|\varphi_{i}\rangle=\alpha_i|\uparrow_{s_i}\rangle+\beta_i|\downarrow_{s_i}\rangle$, where $|\alpha_i|^2+|\beta_i|^2=1$ ($i=1, 2, 3, 4$),  two entangled photons  \emph{A} and \emph{B}, nearly resonant with the transition frequencies $\omega^\uparrow_{A}$ and $\omega^\uparrow_{B}$ of the SiV$^-$ and GeV$^-$ color centers respectively, are generated by an entangled photon source  and serve as a quantum data bus to connect these qubits, enabling the implementation of parallel distributed CNOT gates. These photons are in the state
\begin{align}    
	|\varphi_p'\rangle= \frac{1}{\sqrt{2}}\sum_{m=0}^{1}\mathcal{T}_A(m)\mathcal{T}_B(m)|\varphi_p\rangle, 
	\label{rcoess}
\end{align}
where $|\varphi_p\rangle=( |H_AH_B\rangle+|V_AV_B\rangle)/\sqrt{2}$ is the polarization entangled state. Upon arriving at nodes $\emph{A}$ and $\emph{B}$, photon $\emph{A}$ directly impinges into the CPF {gate}, referred to as $s_1$. Photon $\emph{B}$ undergoes the Hadamard-like transformation $|H\rangle\rightarrow -|A\rangle$ and $|V\rangle\rightarrow |D\rangle$ before impinging into the CPF gate, referred to as $s_2$. The Hadamard transformation is performed on $s_2$ both before and after the scattering process: $\left|\uparrow_{s_2}\right\rangle\rightarrow(\left|\uparrow_{s_2}\right\rangle+\left|\downarrow_{s_2}\right\rangle)/\sqrt{2}$ and 
$\left|\downarrow_{s_2}\right\rangle\rightarrow(\left|\uparrow_{s_2}\right\rangle -\left|\downarrow_{s_2}\right\rangle)/\sqrt{2}$. The combined state of photons $\emph{A}$ and $\emph{B}$ and the four stationary qubits $s_1-s_4$ evolves into
 \begin{align} 	
 |\tilde{\Psi}_1\rangle=&\frac{1}{\sqrt{2\eta_0}}\big\{\big[|H_A\rangle|H_B\rangle( r_{A}^\uparrow\alpha _1|\uparrow_{s_1}\rangle+r_{A}^\downarrow\beta _1|\downarrow_{s_1}\rangle)\nonumber\\
 	&+|V_A\rangle|H_B\rangle \left(\alpha _1|\uparrow_{s_1}\rangle+\beta _1|\downarrow_{s_1}\rangle \right)\big]\big(\mathcal{R}_2^+|\uparrow_{s_2}\rangle\nonumber\\
 	&+\mathcal{R}_2^-|\downarrow_{s_2}\rangle\big)-2\big[|H_A\rangle|V_B\rangle(r_{A}^\uparrow\alpha_1|\uparrow_{s_1}\rangle\nonumber\\
 	&+r_{A}^\downarrow\beta_1|\downarrow_{s_1}\rangle))-|V_A\rangle|V_B\rangle (\alpha_1|\uparrow_{s_1}\rangle+\beta_1|\downarrow_{s_1}\rangle)\big]\nonumber\\
 	&\otimes(\alpha_2|\uparrow_{s_2}\rangle+\beta_2|\downarrow_{s_2}\rangle)\big\},
\end{align} 
where the normalized coefficient is $\eta_0= (|r_A^\uparrow\alpha_1|^2+|r_A^\downarrow\beta_1|^2+1)\big[|r_B^\uparrow|^2+|r_B^\downarrow|^2+(\alpha_2^\ast\beta_2+\alpha_2\beta_2^\ast)(|r_B^\uparrow|^2-|r_B^\downarrow|^2)+2\big]$ and 
{$\mathcal{R}_u^\pm=r_{B}^\uparrow(\alpha_u+\beta_u)\pm r_{B}^\downarrow(\alpha_u-\beta_u)$ for $u=2$ are used for simplicity}.

Subsequently, photons $\emph{A}$ and $\emph{B}$ undergo the Hadamard transformation via a HWP, evolving the combined system state into
\begin{align}
	|\tilde{\Psi}_2\rangle=&\frac{1}{\sqrt{\eta_c}} \mathcal{\hat{T}}\big(|H_A\rangle|H_B\rangle|\Psi_{c_1}\rangle+|H_A\rangle|V_B\rangle|\Psi_{c_2}\rangle\nonumber\\
	&-|V_A\rangle|H_B\rangle|\Psi_{c_3}\rangle -|V_A\rangle|V_B\rangle|\Psi_{c_4}\rangle\big)|\varphi_{s_3}\rangle|\varphi_{s_4}\rangle,\label{stateOne} 
\end{align} 
where four auxiliary states are introduced for simplicity of notation and denote heralded outputs after applying four different distributed quantum gates $\hat{U}_{c_m}$ for $m=1,\ldots, 4$, on qubits $\rm s_1$ and $\rm s_2$. These can be described as follows:
\begin{align}		
	|\Psi_{c_1}\rangle=&\frac{1}{\sqrt{\eta_{c_1}}}\big\{-[r_1^-\alpha _1|\uparrow_{s_1}\rangle+r_2^-\beta _1|\downarrow_{s_1}\rangle]\nonumber\\
	&\otimes\big(2\alpha_2|\uparrow_{s_2}\rangle+2\beta_2|\downarrow_{s_2}\rangle\big)+[r_1^+\alpha _1|\uparrow_{s_1}\rangle\nonumber\\
	&+r_2^+\beta _1|\downarrow_{s_1}\rangle](\mathcal{R}_2^+|\uparrow_{s_2}\rangle+\mathcal{R}_2^-|\downarrow_{s_2}\rangle)\big\},\nonumber\\
	|\Psi_{c_2}\rangle=&\frac{1}{\sqrt{\eta_{c_2}}}\big\{[r_1^-\alpha _1|\uparrow_{s_1}\rangle+r_2^-\beta _1|\downarrow_{s_1}\rangle]\nonumber\\
	&\otimes\big(2\alpha_2|\uparrow_{s_2}\rangle+2\beta_2|\downarrow_{s_2}\rangle\big)+[r_1^+\alpha _1|\uparrow_{s_1}\rangle\nonumber\\
	&+r_2^+\beta _1|\downarrow_{s_1}\rangle](\mathcal{R}_2^+|\uparrow_{s_2}\rangle+\mathcal{R}_2^-|\downarrow_{s_2}\rangle)\big\},\nonumber\\
	|\Psi_{c_3}\rangle=&\frac{1}{\sqrt{\eta_{c_3}}}\big\{[r_1^+\alpha _1|\uparrow_{s_1}\rangle+r_2^+\beta _1|\downarrow_{s_1}\rangle]\nonumber\\
	&\otimes\big(2\alpha_2|\uparrow_{s_2}\rangle+2\beta_2|\downarrow_{s_2}\rangle\big)-[r_1^-\alpha _1|\uparrow_{s_1}\rangle\nonumber\\
	&+r_2^-\beta _1|\downarrow_{s_1}\rangle](\mathcal{R}_2^+|\uparrow_{s_2}\rangle+\mathcal{R}_2^-|\downarrow_{s_2}\rangle\textsc)\big\},\nonumber\\
	|\Psi_{c_4}\rangle=  &\frac{1}{\sqrt{\eta_{c_4}}}\big\{-[r_1^+\alpha _1|\uparrow_{s_1}\rangle+r_2^+\beta _1|\downarrow_{s_1}\rangle]\nonumber\\
	&\big(2\alpha_2|\uparrow_{s_2}\rangle+2\beta_2|\downarrow_{s_2}\rangle\big)-[r_1^-\alpha _1|\uparrow_{s_1}\rangle\nonumber\\
	&+r_2^-\beta _1|\downarrow_{s_1}\rangle](\mathcal{R}_2^+|\uparrow_{s_2}\rangle+\mathcal{R}_2^-|\downarrow_{s_2}\rangle)\big\},
\end{align} 
where $r_1^\pm=r_A^\uparrow\pm1$, $r_2^\pm=r_A^\downarrow\pm1$  $\eta_{c_m}$ for $i=1,\cdots, 4$, are the normalized coefficients, that are summed to form the total coefficient $\eta_{c}=\sum_{m=1}^{4}\eta_{c_m}$ as follows:
\begin{align}
	\eta_{c_1}=&|2 r_1^-\alpha _1\alpha _2-r_1^+\alpha _1\mathcal{R}_2^+|^2+|2 r_1^-\alpha _1\beta _2-r_1^+\alpha _1\mathcal{R}_2^-|^2\nonumber\\
&+|2 r_2^-\beta _1\alpha _2-r_2^+\beta _1\mathcal{R}_2^+|^2+|2 r_2^-\beta _1\beta _2-r_2^+\beta _1\mathcal{R}_2^-|^2,\nonumber\\
\eta_{c_2}=&|2 r_1^-\alpha _1\alpha _2+r_1^+\alpha _1\mathcal{R}_2^+|^2+|2 r_1^-\alpha _1\beta _2+r_1^+\alpha _1\mathcal{R}_2^-|^2\nonumber\\
&+|2 r_2^-\beta _1\alpha _2+r_2^+\beta _1\mathcal{R}_2^+|^2+|2 r_2^-\beta _1\beta _2+r_2^+\beta _1\mathcal{R}_2^-|^2,\nonumber\\
\eta_{c_3}=&|2 r_1^+\alpha _1\alpha _2-r_1^-\alpha _1\mathcal{R}_2^+|^2+|2 r_1^+\alpha _1\beta _2-r_1^-\alpha _1\mathcal{R}_2^-|^2\nonumber\\
&+|2 r_2^+\beta _1\alpha _2-r_2^-\beta _1\mathcal{R}_2^+|^2+|2 r_2^+\beta _1\beta _2-r_2^-\beta _1\mathcal{R}_2^-|^2,\nonumber\\
\eta_{c_4}=&|2 r_1^+\alpha _1\alpha _2+r_1^-\alpha _1\mathcal{R}_2^+|^2+|2 r_1^+\alpha _1\beta _2+r_1^-\alpha _1\mathcal{R}_2^-|^2\nonumber\\
&+|2 r_2^+\beta _1\alpha _2+r_2^-\beta _1\mathcal{R}_2^+|^2+|2 r_2^+\beta _1\beta _2+r_2^-\beta _1\mathcal{R}_2^-|^2.
\end{align}
To decouple qubits $s_1s_2$ from the photon polarization for implementing the distributed CNOT gate on qubits $s_3s_4$, a photon state exchange transformation, $\mathcal{\hat{T}}(1)|H\rangle\leftrightarrow\mathcal{\hat{T}}(0)|V\rangle$, is applied to photons $\emph{A}$ and $\emph{B}$ using the gate $U_{\rm s}$, as described in Appendix D. The combined state evolves into
\begin{align}	
	|\tilde{\Psi}_3\rangle=&|G_{12}\rangle|\varphi_p\rangle|\varphi_{s_3}\rangle|\varphi_{s_4}\rangle,\\
	|G_{12}\rangle=&\frac{1}{\sqrt{\eta_{c}}} [\mathcal{\hat{T}}_A(0)\mathcal{\hat{T}}_B(0)|\Psi_{c_1}\rangle+\mathcal{\hat{T}}_A(0)\mathcal{\hat{T}}_B(1)|\Psi_{c_2}\rangle\nonumber\\ &-\mathcal{\hat{T}}_A(1)\mathcal{\hat{T}}_B(0)|\Psi_{c_3}\rangle-\mathcal{\hat{T}}_A(1)\mathcal{\hat{T}}_B(1)|\Psi_{c_4}\rangle],
\end{align} 
where the polarization state of photons \emph{A} and \emph{B} have been reset to   $|\varphi_p\rangle$, and the time bins are now entangled with qubits $s_1s_2$. 

The photon $\emph{A}$ then interacts with qubit $s_3$ directly, and the photon $\emph{B}$ undergoes the Hadamard-like transformation and then impinges into the CPF {gate}, referred to as $s_4$, before and after which the Hadamard transformation is performed on $\rm s_4$. The combined state of photons $\emph{A}$ and $\emph{B}$ and four stationary qubits $s_1s_2s_3s_4$ evolves into
\begin{align}
	|\tilde{\Psi}_4\rangle=&\frac{1}{\sqrt{2\eta'_{0}}}|G_{12}\rangle\big\{\big[|H_A\rangle|H_B\rangle( r_{A}^\uparrow\alpha _3|\uparrow_{s_3}\rangle+r_{A}^\downarrow\beta _3|\downarrow_{s_3}\rangle)\nonumber\\
	&+|V_A\rangle|H_B\rangle \left(\alpha _3|\uparrow_{s_3}\rangle+\beta _3|\downarrow_{s_3}\rangle \right)\big](\mathcal{R}_4^+|\uparrow_{s_4}\rangle\nonumber\\
	&+\mathcal{R}_4^-|\downarrow_{s_4}\rangle)-\big[|H_A\rangle|V_B\rangle( r_{A}^\uparrow\alpha_3|\uparrow_{s_3}\rangle\nonumber\\
	&+r_{A}^\downarrow\beta_3|\downarrow_{s_3}\rangle)-|V_A\rangle|V_B\rangle \left(\alpha _3|\uparrow_{s_3}\rangle+\beta _3|\downarrow_{s_3}\rangle \right)\big]\nonumber\\
	&\otimes(2\alpha_4|\uparrow_{s_4}\rangle+2\beta_4|\downarrow_{s_4}\rangle)\big\},
	\label{stateTwo}
\end{align} 
where the normalized coefficient is $\eta'_{0} =(|r_A^\uparrow\alpha_3|^2+|r_A^\downarrow\beta_3|^2+1)[|r_B^\uparrow|^2+|r_B^\downarrow|^2+(\alpha_4^\ast\beta_4+\alpha_4\beta_4^\ast)(|r_B^\uparrow|^2-|r_B^\downarrow|^2)+2]$.

To complete the distributed CNOT gate on qubits $\rm s_3$ and $\rm s_4$, the Hadamard transformation is applied to both photons, evolving the combined-system state into
\begin{align}	
	|\tilde{\Psi}_5\rangle=&\frac{1}{\sqrt{\eta'_{c}}}|G_{12}\rangle\big(|H_A\rangle|H_B\rangle |\Psi'_{c_1}\rangle+|H_A\rangle|V_B\rangle|\Psi'_{c_2}\rangle\nonumber\\
	&-|V_A\rangle|H_B\rangle|\Psi'_{c_3}\rangle-|V_A\rangle|V_B\rangle|\Psi'_{c_4}\rangle\big), \label{s10}
\end{align}
where four auxiliary states are introduced for simplicity of {}{notation} and denote, in principle, the heralded outputs after applying four different distributed quantum gates $\hat{U}'_{c_i}$ for $i=1,\ldots, 4$, on the qubits $\rm s_3$ and $\rm s_4$ as follows:
\begin{align}
	|\Psi'_{c_1}\rangle=&\frac{1}{\sqrt{\eta'_{c_1}}}\big\{-[r_1^-\alpha _3|\uparrow_{s_3}\rangle+r_2^-\beta _3|\downarrow_{s_3}\rangle]\nonumber\\
	&\otimes\big(2\alpha_4|\uparrow_{s_4}\rangle+2\beta_4|\downarrow_{s_4}\rangle\big)+[r_1^+\alpha _3|\uparrow_{s_3}\rangle\nonumber\\
	&+r_2^+\beta _3|\downarrow_{s_3}\rangle]\big\{\mathcal{R}_4^+|\uparrow_{s_4}\rangle+\mathcal{R}_4^-|\downarrow_{s_4}\rangle\big\}\big\},\nonumber\\
	|\Psi'_{c_2}\rangle=&\frac{1}{\sqrt{\eta'_{c_2}}}\big\{[r_1^-\alpha _3|\uparrow_{s_3}\rangle+r_2^-\beta _3|\downarrow_{s_3}\rangle]\nonumber\\
	&\otimes\big(2\alpha_4|\uparrow_{s_4}\rangle+2\beta_4|\downarrow_{s_4}\rangle\big)+[r_1^+\alpha _3|\uparrow_{s_3}\rangle\nonumber\\
	&+r_2^+\beta _3|\downarrow_{s_3}\rangle]\big\{\mathcal{R}_4^+|\uparrow_{s_4}\rangle+\mathcal{R}_4^-|\downarrow_{s_4}\rangle\big\}\big\},\nonumber\\
	|\Psi'_{c_3}\rangle=&\frac{1}{\sqrt{\eta'_{c_3}}}\big\{[r_1^+\alpha _3|\uparrow_{s_3}\rangle+r_2^+\beta _3|\downarrow_{s_3}\rangle]\nonumber\\
	&\otimes\big(2\alpha_4|\uparrow_{s_4}\rangle+2\beta_4|\downarrow_{s_4}\rangle\big)-[r_1^-\alpha _3|\uparrow_{s_3}\rangle\nonumber\\
	&+r_2^-\beta _3|\downarrow_{s_3}\rangle]\big\{\mathcal{R}_4^+|\uparrow_{s_4}\rangle+\mathcal{R}_4^-|\downarrow_{s_4}\rangle\big\}\big\},\nonumber\\
	|\Psi'_{c_4}\rangle=  &\frac{1}{\sqrt{\eta'_{c_4}}}\big\{-[r_1^+\alpha _3|\uparrow_{s_3}\rangle+r_2^+\beta _3|\downarrow_{s_3}\rangle]\nonumber\\
	&\otimes\big(2\alpha_4|\uparrow_{s_4}\rangle+2\beta_4|\downarrow_{s_4}\rangle\big)-[r_1^-\alpha _3|\uparrow_{s_3}\rangle\nonumber\\
	&+r_2^-\beta _3|\downarrow_{s_3}\rangle]\big\{\mathcal{R}_4^+|\uparrow_{s_4}\rangle+\mathcal{R}_4^-|\downarrow_{s_4}\rangle\big\}\big\}.
\end{align} 
Here the normalized coefficients  $\eta'_{c_m}$, with the total coefficient $\eta'_{c}=\sum_{m=1}^{4}\eta'_{c_m}$, are described as
\begin{align}
	\eta'_{c_1}=&|2 r_1^-\alpha _3\alpha _4-r_1^+\alpha _3\mathcal{R}_4^+|^2+|2 r_1^-\alpha _3\beta _4-r_1^+\alpha _3\mathcal{R}_4^-|^2\nonumber\\
	&+|2 r_2^-\beta _3\alpha _4-r_2^+\beta _3\mathcal{R}_4^+|^2+|2 r_2^-\beta _3\beta _4-r_2^+\beta _3\mathcal{R}_4^-|^2,\nonumber\\
	\eta'_{c_2}=&|2 r_1^-\alpha _3\alpha _4+r_1^+\alpha _3\mathcal{R}_4^+|^2+|2 r_1^-\alpha _3\beta _4+r_1^+\alpha _3\mathcal{R}_4^-|^2\nonumber\\
	&+|2 r_2^-\beta _3\alpha _4+r_2^+\beta _3\mathcal{R}_4^+|^2+|2 r_2^-\beta _3\beta _4+r_2^+\beta _3\mathcal{R}_4^-|^2,\nonumber\\
	\eta'_{c_3}=&|2 r_1^+\alpha _3\alpha _4-r_1^-\alpha _3\mathcal{R}_4^+|^2+|2 r_1^+\alpha _3\beta _4-r_1^-\alpha _3\mathcal{R}_4^-|^2\nonumber\\
	&+|2 r_2^+\beta _3\alpha _4-r_2^-\beta _3\mathcal{R}_4^+|^2+|2 r_2^+\beta _3\beta _4-r_2^-\beta _3\mathcal{R}_4^-|^2,\nonumber\\
	\eta'_{c_4}=&|2 r_1^+\alpha _3\alpha _4+r_1^-\alpha _3\mathcal{R}_4^+|^2+|2 r_1^+\alpha _3\beta _4+r_1^-\alpha _3\mathcal{R}_4^-|^2\nonumber\\
	&+|2 r_2^+\beta _3\alpha _4+r_2^-\beta _3\mathcal{R}_4^+|^2+|2 r_2^+\beta _3\beta _4+r_2^-\beta _3\mathcal{R}_4^-|^2.
\end{align}

Conditioned on the results of the single-photon measurements on photons $\emph{A}$ and $\emph{B}$, the parallel distributed CNOT gates with two SiV$^-$ qubits $s_1s_2$ as the controls and two GeV$^-$ as the targets can be completed, up to local operations on these qubits. By replacing the practical scattering coefficients $r_{A,B}^\downarrow$ and $r_{A,B}^\uparrow$ with the ideal scattering coefficients $r_{A,B}^\downarrow=-1$ and $r_{A,B}^\uparrow=1$, the output state $|\Psi_2\rangle$ of the combined system, shown in Eq.~(\ref{eq6}) of the main text, can be directly obtained from Eq.~(\ref{s10}), which shows the practical output state of the combined system after applying the parallel distributed CNOT gates on the stationary qubits $s_1s_2$ and $s_3s_4$.

\section{Performance of the parallel distributed CNOT gate}~\label{App.B}

The fidelities of our distributed CNOT gates are defined as the overlap probability between the ideal and practical output states, conditioned on the success of the photon measurements. They, thus, depend on both the qubit states and the photon measurement results, and can be described as follows:
\begin{align}
	F_m=&|\langle \Psi''_m|\Psi_{c_m}\rangle|^2, \\ 
	F_{mn}=&|\langle \Psi''_{mn}|\Psi_{mn}\rangle |^2,
	\label{eqF}
\end{align}
where $|\Psi''_m\rangle= \hat{U}_{c_m}|\varphi_1\rangle|\varphi_2\rangle$ and 
$|\Psi''_{mn}\rangle= \hat{U}_{c_m}\hat{U}'_{c_n}|\varphi_s\rangle$ are the ideal output states of the distributed CNOT gates applied on $s_1s_2$ and on both $s_1s_2$ and $s_3s_4$, respectively, while $|\Psi_{mn}\rangle=|\Psi_{c_m}\rangle|\Psi'_{c_n}\rangle$ is the corresponding realistic output state when both photons $\emph{A}$ and $\emph{B}$ have been measured to be  $\mathcal{\hat{T}}_A(i)\hat{\mathcal{T}}_B(j)\hat{X_A^k}\hat{X_B^l}|HH\rangle$, for $i,j,k,l\in\{0,1\}$. Note that we have incorporated the binary representation of the decimal value with $[m]_{10}=[ij]_2$ and $[n]_{10}=[kl]_2$.

According to Eqs.~(\ref{stateOne}) and (\ref{eqF}), the fidelities of the distributed CNOT gate applied on $s_1s_2$ for four different photon measurement results can be described as
\begin{align}
	F_1	=&\frac{1}{\sqrt{\eta_{c_1}}}\big\{-\alpha_1^\ast\beta_2^\ast\{2 r_1^-\alpha _1\alpha _2+r_1^+\alpha _1\mathcal{R}_2^+\}+\alpha_1^\ast\alpha_2^\ast\nonumber\\
	&\times\{-2 r_1^-\alpha _1\beta _2+r_1^+\alpha _1\mathcal{R}_2^-\}+\beta_1^\ast\alpha_2^\ast\{-2 r_2^-\beta _1\alpha _2\nonumber\\
	&+r_2^+\beta _1\mathcal{R}_2^+\}+\beta_1^\ast\beta_2^\ast\{-2 r_2^-\beta _1\beta _2+r_2^+\beta _1\mathcal{R}_2^-\}\big\},\nonumber\\
	F_2	=&\frac{1}{\sqrt{\eta_{c_2}}}\big\{\alpha_1^\ast\beta_2^\ast\{2 r_1^-\alpha _1\alpha _2+r_1^+\alpha _1\mathcal{R}_2^+\}+\alpha_1^\ast\alpha_2^\ast\nonumber\\
	&\times\{2 r_1^-\alpha _1\beta _2+r_1^+\alpha _1\mathcal{R}_2^-\}-\beta_1^\ast\alpha_2^\ast\{2 r_2^-\beta _1\alpha _2\nonumber\\
	&+r_2^+\beta _1\mathcal{R}_2^+\}-\beta_1^\ast\beta_2^\ast\{2 r_2^-\beta _1\beta _2+r_2^+\beta _1\mathcal{R}_2^-\}\big\},\nonumber\\
	F_3	=&\frac{1}{\sqrt{\eta_{c_3}}}\big\{\alpha_1^\ast\alpha_2^\ast\{2 r_1^+\alpha _1\alpha _2-r_1^-\alpha _1\mathcal{R}_2^+\}+\alpha_1^\ast\beta_2^\ast\nonumber\\
	&\times\{2 r_1^+\alpha _1\beta _2-r_1^-\alpha _1\mathcal{R}_2^-\}+\beta_1^\ast\beta_2^\ast\{2 r_2^+\beta _1\alpha _2\nonumber\\
	&-r_2^-\beta _1\mathcal{R}_2^+\}+\beta_1^\ast\alpha_2^\ast\{2 r_2^+\beta _1\beta _2-r_2^-\beta _1\mathcal{R}_2^-\}\big\},\nonumber\\
	F_4  =&\frac{1}{\sqrt{\eta_{c_4}}}\big\{-\alpha_1^\ast\alpha_2^\ast\{-2 r_1^+\alpha _1\alpha _2-r_1^-\alpha _1\mathcal{R}_2^+\}-\alpha_1^\ast\beta_2^\ast\nonumber\\
	&\times\{-2 r_1^+\alpha _1\beta _2-r_1^-\alpha _1\mathcal{R}_2^-\}+\beta_1^\ast\beta_2^\ast\{-2 r_2^+\beta _1\alpha _2\nonumber\\
	&-r_2^-\beta _1\mathcal{R}_2^+\}+\beta_1^\ast\alpha_2^\ast\{-2 r_2^+\beta _1\beta _2-r_2^-\beta _1\mathcal{R}_2^-\}\big\}.
\end{align}

\begin{figure}[t!]
	\includegraphics[width=8.5cm]{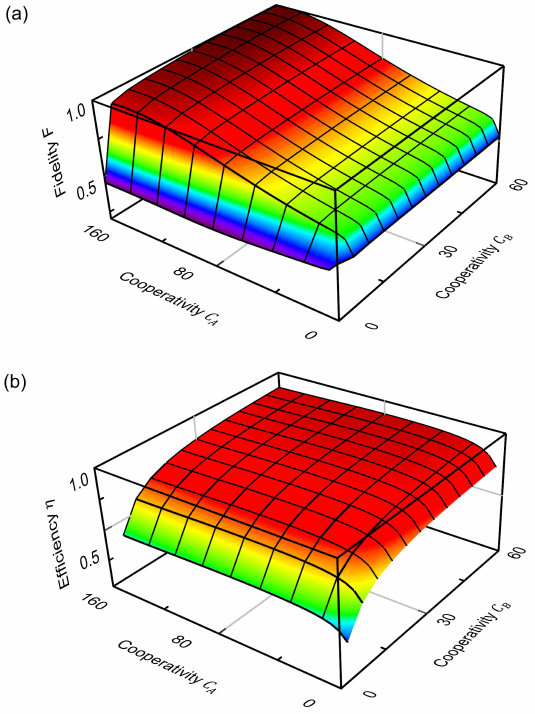}
	\caption{(a) Average fidelity and efficiency of the distributed CNOT gate on the spatially separated SiV$^-$~($s_1$) and GeV$^-$~($s_2$) color centers. Here the normalized parameters are $\Delta_{\uparrow_A}=\Delta_{\uparrow_B}=0$, $\Delta_{\downarrow_A}=\Delta_{\downarrow_B}=100$, $\Delta_{C_A}=1.5$, and $\Delta_{C_B}=0.5$.}\label{figs1} 
\end{figure}

Meanwhile, the corresponding normalized coefficients $\eta_{c_m}$ for $m=1, \ldots, 4$, can be referred to as the efficiencies of the distributed CNOT gate applied on $s_1s_2$ for four different photon measurement results. The total efficiency can then be defined as 
\begin{align}\label{eff}
	\tilde{\eta}_s=\sum_{m=1}^{4}\eta_{c_m}. 
\end{align}

The average fidelity and efficiency for four different photon measurement results and qubit states can, thus, be described as follows:
\begin{align}
	\tilde{F}=&\frac{1}{4}\sum_{m=1}^{4}\int_{0}^{1}\int_{0}^{1}{}{F_m\;} d\alpha_1d\alpha_2,\\
	\tilde{\eta} &=\sum_{m=1}^{4}\int_{0}^{1}\int_{0}^{1}\eta_{c_m}d\alpha_1d\alpha_2,
\end{align}
which are shown, as functions of the color center-cavity cooperativities $C_A$ and $C_B$, in Fig.~\ref{figs1}. The average fidelity reaches approximately $\tilde{F}\simeq0.999$, while the average efficiency is around  $\tilde{\eta}\simeq0.943$ for $C_A=150$ and $C_B=50$, using the normalized parameters $\Delta_{\uparrow_A}=\Delta_{\uparrow_B}=0$, $\Delta_{\downarrow_A}=\Delta_{\downarrow_B}=100$, $\Delta_{C_A}=1.5$, and $\Delta_{C_B}=0.5$.

The fidelities of the parallel distributed CNOT gate applied on qubits $s_1s_2$ for $s_3s_4$ for $16$ different photon measurement results can be defined similarly to the previous description, because the two parallel CNOT gates applied on the qubits $s_1s_2$ and $s_3s_4$ are separable: the former is determined by the time bins and the latter is determined by the polarizations of photons $\emph{A}$ and $\emph{B}$, as shown in Eq.~(\ref{stateTwo}). The average fidelity and efficiency across the $16$ photon measurement results and the qubit states can, thus, be described as:
\begin{align}
	F=&\frac{1}{16}\sum_{m,n}\int_{0}^{1}\int_{0}^{1}\int_{0}^{1}\int_{0}^{1}F_mF_n'd\alpha_1d\alpha_2d\alpha_3d\alpha_4,\\
	\eta &=\sum_{m,n}\int_{0}^{1}\int_{0}^{1}\int_{0}^{1}\int_{0}^{1}\eta_{c_m} \eta'_{c_n}d\alpha_1d\alpha_2d\alpha_3d\alpha_4.
\end{align}
Here $F_m'$ ($m=1, \ldots, 4$) is the fidelity of the distributed CNOT gate applied on qubits $s_3s_4$ and can be described as 
\begin{align}
	F_1'
	=&\frac{1}{\sqrt{\eta'_{c_1}}}\big\{-\alpha_3^\ast\beta_4^\ast\{2 r_1^-\alpha _3\alpha _4+r_1^+\alpha _3\mathcal{R}_4^+\}+\alpha_3^\ast\alpha_4^\ast\nonumber\\
	&\times\{-2 r_1^-\alpha _3\beta _4+r_1^+\alpha _3\mathcal{R}_4^-\}+\beta_3^\ast\alpha_4^\ast\{-2 r_2^-\beta _3\alpha _4\nonumber\\
	&+r_2^+\beta _3\mathcal{R}_4^+\}+\beta_3^\ast\beta_4^\ast\{-2 r_2^-\beta _3\beta _4+r_2^+\beta _3\mathcal{R}_4^-\}\big\},\nonumber\\
	F_2'=&\frac{1}{\sqrt{\eta'_{c_2}}}\big\{\alpha_3^\ast\beta_4^\ast\{2 r_1^-\alpha _3\alpha _4+r_1^+\alpha _3\mathcal{R}_4^+\}+\alpha_3^\ast\alpha_4^\ast\nonumber\\
	&\times\{2 r_1^-\alpha _3\beta _4+r_1^+\alpha _3\mathcal{R}_4^-\}-\beta_3^\ast\alpha_4^\ast\{2 r_2^-\beta _3\alpha _4\nonumber\\
	&+r_2^+\beta _3\mathcal{R}_4^+\}-\beta_3^\ast\beta_4^\ast\{2 r_2^-\beta _3\beta _4+r_2^+\beta _3\mathcal{R}_4^-\}\big\},\nonumber\\
	F_3'=&\frac{1}{\sqrt{\eta'_{c_3}}}\big\{\alpha_3^\ast\alpha_4^\ast\{2 r_1^+\alpha _3\alpha _4-r_1^-\alpha _3\mathcal{R}_4^+\}+\alpha_3^\ast\beta_4^\ast\nonumber\\
	&\times\{2 r_1^+\alpha _3\beta _4-r_1^-\alpha _3\mathcal{R}_4^-\}+\beta_3^\ast\beta_4^\ast\{2 r_2^+\beta _3\alpha _4\nonumber\\
	&-r_2^-\beta _3\mathcal{R}_4^+\}+\beta_3^\ast\alpha_4^\ast\{2 r_2^+\beta _3\beta _4-r_2^-\beta _3\mathcal{R}_4^-\}\big\},\nonumber\\
	F_4'=&\frac{1}{\sqrt{\eta'_{c_4}}}\big\{-\alpha_3^\ast\alpha_4^\ast\{-2 r_1^+\alpha _3\alpha _4-r_1^-\alpha _3\mathcal{R}_4^+\}-\alpha_3^\ast\beta_4^\ast\nonumber\\
	&\times\{-2 r_1^+\alpha _3\beta _4-r_1^-\alpha _3\mathcal{R}_4^-\}+\beta_3^\ast\beta_4^\ast\{-2 r_2^+\beta _3\alpha _4\nonumber\\
	&-r_2^-\beta _3\mathcal{R}_4^+\}+\beta_3^\ast\alpha_4^\ast\{-2 r_2^+\beta _3\beta _4-r_2^-\beta _3\mathcal{R}_4^-\}\big\}.
\end{align}

\section{{}{Influence of imperfect entangled photon pairs on the distributed CNOT gate}}~\label{App.C}
{}{We now assume that the imperfect photon pair, used for implementing a distributed CNOT gate, is in the mixed entangled state
\begin{align}
	\rho_0=F_0|\varphi_p\rangle\langle\varphi_p|+(1-F_0)|\varphi^-_p\rangle\langle\varphi^-_p|, 
\end{align}
where the desired state $|\varphi_p\rangle$ occurs with probability $F_0$, which equals the entanglement fidelity of $\rho_0$. The state $|\varphi^-_p\rangle = (|H_A H_B\rangle - |V_A V_B\rangle)/\sqrt{2}$ represents an error and occurs with probability $(1 - F_0)$}.

The state of the combined system, comprising a photon pair and stationary qubits $s_1$ and $s_2$, evolves independently for the photon pairs in the states $|\varphi_p\rangle$ and $|\varphi^-_p\rangle$. The final state of the combined system, prior to directing the photon pair to the corresponding measurement units, can be described as:
\begin{align}
	\rho^c_0=&F_0|\Phi_2\rangle\langle\Phi_2|+(1-F_0)|\Phi^-_2\rangle\langle\Phi_2^-|,\nonumber\\
	|\Phi_2\rangle=& \frac{1}{2}\Big(\hat{U}_{c_1}|HH\rangle+\hat{U}_{c_2}|HV\rangle-\hat{U}_{c_3}|VH\rangle -\hat{U}_{c_4}|VV\rangle\Big)\nonumber\\
	&\otimes|\varphi_1\rangle|\varphi_2\rangle, \nonumber\\
	|\Phi^-_2\rangle=&\frac{1}{2}\Big(\hat{U}_{c_1}|VH\rangle+\hat{U}_{c_2}|VV\rangle-\hat{U}_{c_3}|HH\rangle -\hat{U}_{c_4}|HV\rangle\Big)\nonumber\\
	&\otimes|\varphi_1\rangle|\varphi_2\rangle, 
\end{align}
where 
\begin{align}
	\hat{U}_{c_1}|\varphi_1\rangle|\varphi_2\rangle=\alpha_1\hat{X}_2|\uparrow\rangle |\varphi_2\rangle+\beta_1\hat{I}_2|\downarrow\rangle|\varphi_2\rangle
\end{align}
 applies the desired CNOT gate on the spatially separated stationary qubits $s_1$ and $s_2$, $\hat{U}_{c_2}=\hat{Z}_1\hat{U}_{c_1}$, $\hat{U}_{c_3}=\hat{X}_2\hat{U}_{c_1}$, and $\hat{U}_{c_4}=-\hat{Z}_1\hat{X}_2\hat{U}_{c_1}$.

The two polarized photons are then detected simultaneously. The state of the stationary qubits $s_1$ and $s_2$ is projected onto the following states:
\begin{align}
	\rho_1^s &= f_{13}, &
	\rho_2^s &= f_{24}, \nonumber\\
	\rho_3^s &= f_{31}, &
	\rho_4^s &= f_{42},
	\label{C3}
\end{align}
where
\begin{align}
	f_{xy} =& F_0 \hat{U}_{c_x} \, |\varphi_1\rangle |\varphi_2\rangle \langle \varphi_1| \langle \varphi_2| \, \hat{U}_{c_x}^\dagger\nonumber\\
&	+ (1 - F_0) \, \hat{U}_{c_y} \, |\varphi_1\rangle |\varphi_2\rangle \langle \varphi_1| \langle \varphi_2| \, \hat{U}_{c_y}^\dagger
	\label{C3a}
\end{align}
for the measurement results  $\{|HH\rangle$, $|HV\rangle$, $|VH\rangle$, and $|VV\rangle\}$, respectively. For a perfect photon pair with unity fidelity $F_0=1$, the states $\rho^s_i$ ($i=1,\dots,4$) correspond to the ideal cases for performing the distributed dual-species CNOT gate. For imperfect photon pairs with $F_0<1$, errors proportional to $(1 - F_0)$ are introduced.  In principle,  these errors can be suppressed by employing an entanglement purification protocol~\cite{Yan2023Advances} before the photons interact with the spins. After a single round of this entanglement purification protocol, the entanglement fidelity of the photon pair improves to
be $F'_0=F^2_0/[F_0^2+(1-F_0)^2]$, leading to $F'_0>0.999$ for $F_0>0.969$.

\section{ Parallel distributed CNOT gate for superconducting and Si$\rm V^-$ qubits}~\label{App.D}

\begin{figure*}[th!]
	\includegraphics[width=17.2 cm]{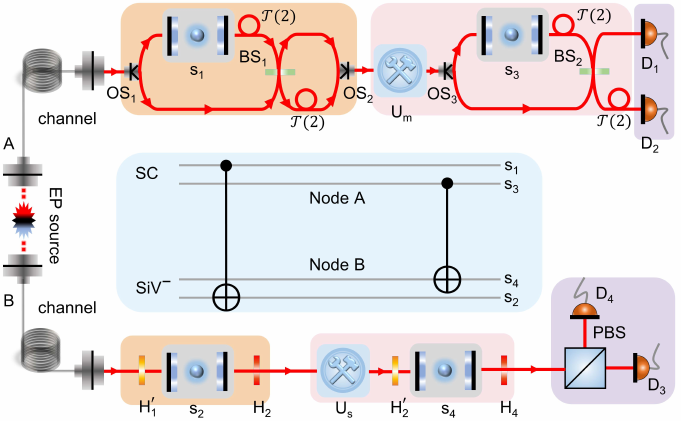}
	\caption{Schematic of the parallel distributed CNOT gate with two  superconducting qubits $\rm s_{1}$ and $\rm s_{3}$ as the controls and two $\rm SiV^-$ qubits $\rm s_{2}$ and $\rm s_{4}$ as the corresponding targets. BS$_i$ denotes a balanced microwave beam splitter and combines with time-delay $\mathcal{T}(2)$ to complete the Hadamard transformation for microwave photons. H$_i$ is a $\pi/8$-half-wave plate and performs the Hadamard transformation. H$_i'$ performs a Hadamard-like transformation. PBS is a polarizing	beam splitter that transmits (reflects) $H$-polarized~($V$-polarized) photons.} \label{figs2}
\end{figure*}

In this section, we demonstrate that our protocol can implement parallel distributed CNOT gates on spatially separated superconducting (SC) qubits and color center qubits (i.e., Si$\rm V^-$ electron spins), which operate in the microwave and optical regimes, respectively. For the Si$\rm V^-$ qubit, its coupling to an optical cavity enables a CPF gate, facilitating effective interaction with an optical photon. For an SC qubit, its dispersive interaction with a microwave resonator results in a controlled phase (CPHASE) gate for an input microwave photon, as described by the following transformation~\cite{Kono2018Quantum}:
\begin{align}
	|1\rangle|g\rangle &\rightarrow |1\rangle|g\rangle, \qquad 
	|1\rangle|e\rangle \rightarrow -|1\rangle|e\rangle, 
\end{align}
where $|1\rangle$ represents the single-photon Fock state, and $|g\rangle$ and $|e\rangle$ are the ground and excited states of the superconducting qubit. In practice, this CPHASE gate can be used to implement a CNOT gate between a time-bin-encoded microwave photon and an SC qubit, which can then be used to implement distributed CNOT gates on spatially separated SC and Si$\rm V^-$ qubits, as shown in Fig.~\ref{figs2}.

Consider now two SC qubits, each dispersively coupled to a microwave resonator in node A, initialized in the states $|\phi_{s_1}\rangle = (\alpha_1|g_{s_1}\rangle + \beta_1|e_{s_1}\rangle)/\sqrt{2}$ and $|\phi_{s_3}\rangle = (\alpha_3|g_{s_3}\rangle + \beta_3|e_{s_3}\rangle)/\sqrt{2}$. Two Si$\rm V^-$ qubits embedded in CPF gates in node B are initialized as {$\left|\phi_{s_2}\right\rangle = (\alpha_2\left|\uparrow_{s_2}\right\rangle + \beta_2\left|\downarrow_{s_2}\right\rangle)/\sqrt{2}$} and $\left|\phi_{s_4}\right\rangle = (\alpha_4\left|\uparrow_{s_4}\right\rangle + \beta_4\left|\downarrow_{s_4}\right\rangle)/\sqrt{2}$. An auxiliary entangled-photon source generates a pair of hybrid entangled photons in the state
	\begin{align}
		|\phi_p\rangle = \frac{1}{2} \sum_{i=0}^{1} 
		\big[ \mathcal{\hat{T}}_A(i)\mathcal{\hat{T}}_B(i) + 
		\mathcal{\hat{T}}_A(i+2)\mathcal{\hat{T}}_B(i)\hat{X}_p \big]
		|1_A\rangle|H_B\rangle,
	\end{align}
where photon A (B) is in the microwave (optical) regime and nearly resonant with the  SC qubit~(SiV$^-$) transition. $\hat{X}_p$ flips the polarization of photon B, and $\mathcal{\hat{T}}_A(i)$ [$\mathcal{\hat{T}}_B(i)$] introduces a time delay for photon A (B).

The microwave photon, in the states $\mathcal{\hat{T}}(0)|1\rangle$ and $\mathcal{\hat{T}}(1)|1\rangle$, is directed into the upper path and interacts with qubit $s_1$ via OS$_1$, while states $\mathcal{\hat{T}}(2)|1\rangle$ and $\mathcal{\hat{T}}(3)|1\rangle$ are directed into the lower path. After the interaction, the composite system evolves into
	\begin{align}
		|\xi_1\rangle = \frac{1}{\sqrt{2}}\mathcal{\hat{T}}
		\big[ \mathcal{\hat{T}}_A(0)\hat{Z}'_1|1_A\rangle |H_B\rangle 
		+ \mathcal{\hat{T}}_A(2)|1_A\rangle|V_B\rangle \big]|\varphi_{\tilde{s}}\rangle, 
	\end{align}
where $\hat{Z}'_1 = |g\rangle_{s_1}\langle g| - |e\rangle_{s_1}\langle e|$ introduces a phase $\pi$ to the SC qubit $s_1$, and $|\varphi_{\tilde{s}}\rangle = |\phi_{s_1}\rangle|\phi_{s_2}\rangle|\phi_{s_3}\rangle|\phi_{s_4}\rangle$. The operator $\mathcal{\hat{T}} = \big( \mathcal{\hat{T}}_A(0)\mathcal{\hat{T}}_B(0) + \mathcal{\hat{T}}_A(1)\mathcal{\hat{T}}_B(1) \big)/\sqrt{2}$ denotes the combined time delays.

Meanwhile, photon B passes through $H_1'$ and interacts with the SiV$^-$ qubit $s_2$ before and after the Hadamard transformation of $s_2$. The combined state of the composite system then evolves into
	\begin{align}
		|\xi_2\rangle =& \frac{1}{2\sqrt{2}} \mathcal{\hat{T}}\big[\mathcal{\hat{T}}_A(0)\hat{Z}'_1\hat{X}_2|1_A\rangle|H_B\rangle
		- \mathcal{\hat{T}}_A(0)\hat{Z}'_1|1_A\rangle|V_B\rangle\nonumber\\
		&+ \mathcal{\hat{T}}_A(2)\hat{X}_2|1_A\rangle|H_B\rangle + \mathcal{\hat{T}}_A(2)|1_A\rangle|V_B\rangle\big]|\varphi_{\tilde{s}}\rangle, 
	\end{align} 
where the microwave and optical photons AB entangle with the SC qubit $s_1$ and the SiV$^-$ qubit $s_2$ in a hybrid way.

The Hadamard transformation of the time-bin-encoded microwave photon, $\mathcal{\hat{T}}(i) \rightarrow [\mathcal{\hat{T}}(i) + \mathcal{\hat{T}}(i+2)]/\sqrt{2}$ and $\mathcal{\hat{T}}(i+2) \rightarrow [\mathcal{\hat{T}}(i) - \mathcal{\hat{T}}(i+2)]/\sqrt{2}$ for $i=0,~1$, is applied along with the Hadamard transformation of the polarization-encoded optical photon. The former is completed by introducing a time delay $\mathcal{\hat{T}}(2)$ in the upper path and combining it with that in the lower path at a BS followed by a time delay $\mathcal{\hat{T}}(2)$ in its lower outport. The latter is completed by passing photon B through H$_2$. The combined state of the system after these transformations is
\begin{align}
	|\xi_3\rangle = & \frac{1}{2}\mathcal{\hat{T}}\big[\mathcal{\hat{T}}_A(0)\hat{U}_{\tilde{c}_1}|1_A\rangle|H_B\rangle
	+ \mathcal{\hat{T}}_A(0)\hat{U}_{\tilde{c}_2}|1_A\rangle|V_B\rangle \nonumber\\ 
	&- \mathcal{\hat{T}}_A(2)\hat{U}_{\tilde{c}_3}|1_A\rangle|H_B\rangle
	- \mathcal{\hat{T}}_A(2)\hat{U}_{\tilde{c}_4}|1_A\rangle|V_B\rangle\big]|\varphi_{\tilde{s}}\rangle,  
\end{align} 
where we denote $\hat{U}_{\tilde{c}_1}|\phi_{s_1}\rangle|\phi_{s_2}\rangle = \alpha_1\hat{X}_2|g_{s_1}\rangle|\phi_{s_2}\rangle + \beta_1\hat{I}_2|e_{s_1}\rangle|\phi_{s_2}\rangle$, $\hat{U}_{\tilde{c}_2} = \hat{Z}'_1\hat{U}_{\tilde{c}_1}$, $\hat{U}_{\tilde{c}_3} = \hat{X}_2\hat{U}_{\tilde{c}_1}$, and {$\hat{U}_{\tilde{c}_4} = -\hat{Z}'_1\hat{X}_2\hat{U}_{\tilde{c}_1}$}. We refer to $\hat{U}_{\tilde{c}_1}$ as the target CNOT gate on the SC qubit $s_1$ and the SiV$^-$ qubit $s_3$, and single-qubit operations can complete the target CNOT gate for the other three cases.

Then the two photons pass through ${\rm U}_m$ and ${\rm U}_s$, which perform the operation $\mathcal{\hat{T}}(1)|1\rangle \leftrightarrow \mathcal{\hat{T}}(2)|1\rangle$ and $\mathcal{\hat{T}}(1)|H\rangle \leftrightarrow \mathcal{\hat{T}}(0)|V\rangle$, as shown in Appendix~\ref{App.E}. The combined state evolves into
	\begin{align}
		|\xi_4\rangle =&  \frac{1}{2}\big[\mathcal{\hat{T}}_A(0)\mathcal{\hat{T}}_B(0)\hat{U}_{\tilde{c}_1}
		+ \mathcal{\hat{T}}_A(0)\mathcal{\hat{T}}_B(1) \hat{U}_{\tilde{c}_2}\nonumber\\
		& - \mathcal{\hat{T}}_A(1)\mathcal{\hat{T}}_B(0)\hat{U}_{\tilde{c}_3} 
		- \mathcal{\hat{T}}_A(1)\mathcal{\hat{T}}_B(1) \hat{U}_{\tilde{c}_4}\big]|\varphi_{\tilde{p}}\rangle|\varphi_{\tilde{s}}\rangle, 
	\end{align} 
where the entanglement between the polarization of optical photon \emph{B} and the time-bin mode of microwave photon \emph{A} has been reset to $|\varphi_{\tilde{p}}\rangle = [\mathcal{\hat{T}}_A(0)\mathcal{\hat{T}}_B(0) + \mathcal{\hat{T}}_A(2)\mathcal{\hat{T}}_B(0)\hat{X}_p]|1_A\rangle|H_B\rangle/\sqrt{2}$.

We can repeat the above process to perform the CNOT gate on the SC qubit $s_2$ and the SiV$^-$ qubit $s_4$.After microwave photon A passes through $\rm{OS}_3$, which directs it in the same way as $\rm{OS}_1$, photon \emph{A} interacts with qubit $s_3$ and evolves the composite system into
	\begin{align}
		|\xi_5\rangle = & \frac{1}{\sqrt{2}}\big[\mathcal{\hat{T}}_A(0)\hat{Z}'_3|1_A\rangle|H_B\rangle  
		+\mathcal{\hat{T}}_A(2){|1_A\rangle|V_B\rangle }\big]\nonumber\\
		&\otimes|\phi_{s_3}\rangle|\phi_{s_4}\rangle|G_{12}'\rangle,  \\
		|G_{12}'\rangle = & \frac{1}{2}\big[\mathcal{\hat{T}}_A(0)\mathcal{\hat{T}}_B(0)\hat{U}_{\tilde{c}_1}
		+ \mathcal{\hat{T}}_A(0)\mathcal{\hat{T}}_B(1) \hat{U}_{\tilde{c}_2}\nonumber\\
		& - \mathcal{\hat{T}}_A(1)\mathcal{\hat{T}}_B(0)\hat{U}_{\tilde{c}_3} 
		- \mathcal{\hat{T}}_A(1)\mathcal{\hat{T}}_B(1) \hat{U}_{\tilde{c}_4}\big]|\phi_{s_1}\rangle|\phi_{s_2}\rangle, 
	\end{align} 
where $\hat{Z}'_3 = |g\rangle_{s_3}\langle g| - |e\rangle_{s_3}\langle e|$ introduces a phase $\pi$ for the SC qubit $s_3$.We note that a time delay $\mathcal{\hat{T}}_A(i)\mathcal{\hat{T}}_B(j)$ has been explicitly incorporated to distinguish four single-qubit operations required to complete the CNOT gate on the SC qubit $s_1$ and the SiV$^-$ qubit $s_3$.

Meanwhile, photon \emph{B} passes through $H_2'$ and interacts with qubit $s_4$ before and after the Hadamard transformation of $s_4$. The combined states of the composite system evolve into
	\begin{align}
		|\xi_{6}\rangle = & \frac{1}{2\sqrt{2}}\big[
		\mathcal{\hat{T}}_A(0)\hat{Z}'_3\hat{X}_4|1_A\rangle|H_B\rangle
		- \mathcal{\hat{T}}_A(0)\hat{Z}'_3|1_A\rangle|V_B\rangle  \nonumber\\
		&+ \mathcal{\hat{T}}_A(2)\hat{X}_4|1_A\rangle|H_B\rangle
		+ \mathcal{\hat{T}}_A(2)|1_A\rangle|V_B\rangle\big]\nonumber\\
		&\otimes|\phi_{s_3}\rangle|\phi_{s_4}\rangle|G_{12}'\rangle. 
	\end{align} 
To complete the CNOT gate on qubits $s_3$ and $s_4$, the Hadamard transformation of the time-bin state of microwave photon \emph{A} and that of the polarized photon \emph{B} are applied. The combined state of the composite system evolves into
	\begin{align}
		|\xi_{7}\rangle = &  \frac{1}{2}\big[
		\mathcal{\hat{T}}_A(0)\hat{U}'_{\tilde{c}_1}|1_A\rangle|H_B\rangle  
		+ \mathcal{\hat{T}}_A(0)\hat{U}'_{\tilde{c}_2}|1_A\rangle|V_B\rangle  \nonumber\\
		&- \mathcal{\hat{T}}_A(2)\hat{U}'_{\tilde{c}_3}|1_A\rangle|H_B\rangle  
		- \mathcal{\hat{T}}_A(2)\hat{U}'_{\tilde{c}_4}|1_A\rangle|V_B\rangle\big]\nonumber\\
		&\otimes|\phi_{s_3}\rangle|\phi_{s_4}\rangle|G_{12}'\rangle, 
	\end{align} 
where we denote $\hat{U}'_{\tilde{c}_1}|\phi_{s_3}\rangle|\phi_{s_4}\rangle = \alpha_3\hat{X}_4|g_{s_3}\rangle|\phi_{s_4}\rangle + \beta_3\hat{I}_4|e_{s_3}\rangle|\phi_{s_4}\rangle$, $\hat{U}'_{\tilde{c}_2} = \hat{Z}'_3\hat{U}'_{\tilde{c}_1}$, $\hat{U}'_{\tilde{c}_3} = \hat{X}_4\hat{U}'_{\tilde{c}_1}$, and {$\hat{U}'_{\tilde{c}_4} = -\hat{Z}'_3\hat{X}_4\hat{U}'_{\tilde{c}_1}$}. 
The parallel distributed CNOT gates $\hat{U}_{\tilde{c}_1}\hat{U}'_{\tilde{c}_1}$ on two SC qubits and two $\rm{SiV}^-$ qubits can thus be completed in a heralded way, up to local single-qubit operations that depend on the measurement results of microwave and optical single-photon detections.

\section{Schematics of gates U$_s$ and U$_m$}~\label{App.E}

The schematic of the {gate} U$_m$ consists of time delays $\mathcal{\hat{T}}(1)$, optical switches~($\rm OS_1$ {and} $\rm OS_2$), 
polarizing beam splitters ($\rm PBSs$ {and} $\rm PBS's$), and HWPs, as shown in Fig.~\ref{fig1}(c). The  HWP   performs a bit-flip operation on the photon that passes through it, flipping its polarization. $\rm OS_1$ directs a photon in  time-bin state $\mathcal{\hat{T}}(0)$~[$\mathcal{\hat{T}}(1)$] to the upper~(lower) path, while $\rm OS_2$ combines photons from two paths with different time bins into a single path. The $\rm PBSs$  {transmit the} $H$-polarized photons and {reflect the} $V$-polarized photons, while {the} $\rm PBS's$ {transmit the}
$V$-polarized photons and {reflect the} $H$-polarized photons. Therefore, the state evolution of a photon  input into the {gate} U$_s$ in {the} four different polarization and time-bin states can be described as follows:
\begin{align}	\mathcal{\hat{T}}(0)|H\rangle\rightarrow\mathcal{\hat{T}}(1)|H\rangle\rightarrow\mathcal{\hat{T}}(1)|H\rangle\rightarrow\mathcal{\hat{T}}(1)|H\rangle\rightarrow\mathcal{\hat{T}}(0)|H\rangle,\nonumber\\
	\mathcal{\hat{T}}(0)|V\rangle\rightarrow\mathcal{\hat{T}}(1)|V\rangle\rightarrow\mathcal{\hat{T}}(1)|H\rangle\rightarrow\mathcal{\hat{T}}(2)|H\rangle\rightarrow\mathcal{\hat{T}}(1)|H\rangle,\nonumber\\
	\mathcal{\hat{T}}(1)|H\rangle\rightarrow\mathcal{\hat{T}}(1)|H\rangle\rightarrow\mathcal{\hat{T}}(1)|V\rangle\rightarrow\mathcal{\hat{T}}(1)|V\rangle\rightarrow\mathcal{\hat{T}}(0)|V\rangle,\nonumber\\
	\mathcal{\hat{T}}(1)|V\rangle\rightarrow\mathcal{\hat{T}}(1)|V\rangle\rightarrow\mathcal{\hat{T}}(1)|V\rangle\rightarrow\mathcal{\hat{T}}(2)|V\rangle\rightarrow\mathcal{\hat{T}}(1)|V\rangle,
\end{align}
where we have referred to the subspaces \{$\mathcal{\hat{T}}(1)|H\rangle$, $\mathcal{\hat{T}}(2)|H\rangle$, $\mathcal{\hat{T}}(1)|V\rangle$, $\mathcal{\hat{T}}(2)|V\rangle$\} as \{$\mathcal{\hat{T}}(0)|H\rangle$, $\mathcal{\hat{T}}(1)|H\rangle$, $\mathcal{\hat{T}}(0)|V\rangle$, $\mathcal{\hat{T}}(1)|V\rangle$\} after removing a time delay of $\mathcal{\hat{T}}(1)$ for all time bins for simplicity. Therefore, the {gate} U$_s$ performs the state exchange 
\begin{align}	
\mathcal{\hat{T}}(0)|V\rangle\leftrightarrow\mathcal{\hat{T}}(1)|H\rangle, 
\end{align}
for each single photon and decouples the previous qubit pair $s_1s_2$ from the photon polarization, enabling the implementation of the subsequent distributed CNOT gate on the next qubit pair $s_3s_4$.

\begin{figure}[t!]
	\includegraphics[width=8.8cm]{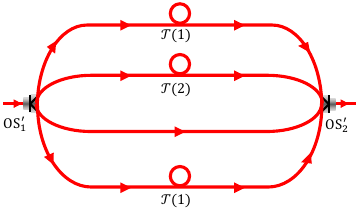}
	{\caption{Schematic  of the gate U$_m$. $\rm OS_1'$ directs the microwave photon with time delays $\mathcal{\hat{T}}(0)$, $\mathcal{\hat{T}}(1)$, $\mathcal{\hat{T}}(2)$, and $\mathcal{\hat{T}}(3)$ into four different paths from top to bottom, respectively. $\rm OS_2'$ combines four paths into one spatial mode.}\label{fig3s}}
\end{figure}

The schematic of the {gate} U$_m$ consists of time delays $\mathcal{\hat{T}}(1)$ and $\mathcal{\hat{T}}(2)$, and optical switches~($\rm OS'_1$ {and} $\rm OS'_2$), as shown in Fig.~\ref{fig3s}.
The optical switch $\rm OS_1'$ directs microwave photons with time delays $\mathcal{\hat{T}}(0)$, $\mathcal{\hat{T}}(1)$, $\mathcal{\hat{T}}(2)$, and $\mathcal{\hat{T}}(3)$ into different paths from top to bottom, respectively. The optical switch $\rm OS_2'$ combines microwave photons in four different time-bin states into one spatial mode. The states of a time-bin encoded microwave photon evolve as follows: 
	\begin{align} \mathcal{\hat{T}}(0)|1\rangle &\rightarrow \mathcal{\hat{T}}(1)|1\rangle \rightarrow \mathcal{\hat{T}}(0)|1\rangle, \notag \\ 
		\mathcal{\hat{T}}(1)|1\rangle &\rightarrow \mathcal{\hat{T}}(3)|1\rangle \rightarrow \mathcal{\hat{T}}(2)|1\rangle, \notag \\
		\mathcal{\hat{T}}(2)|1\rangle &\rightarrow \mathcal{\hat{T}}(2)|1\rangle \rightarrow \mathcal{\hat{T}}(1)|1\rangle, \notag \\ 
		\mathcal{\hat{T}}(3)|1\rangle &\rightarrow \mathcal{\hat{T}}(4)|1\rangle \rightarrow \mathcal{\hat{T}}(3)|1\rangle. 
	\end{align}
Similarly, the subspace $\{\mathcal{\hat{T}}(1)|1\rangle$, $\mathcal{\hat{T}}(3)|1\rangle$, $\mathcal{\hat{T}}(2)|1\rangle$, $\mathcal{\hat{T}}(4)|1\rangle\}$ can be mapped to $\{\mathcal{\hat{T}}(0)|1\rangle$, $\mathcal{\hat{T}}(2)|1\rangle$, $\mathcal{\hat{T}}(1)|1\rangle$, $\mathcal{\hat{T}}(3)|1\rangle\}$ after removing a global delay $\mathcal{\hat{T}}(1)$ for all time bins. Hence, the {gate} U$_m$ implements the transformation of $\mathcal{\hat{T}}(1)|1\rangle \leftrightarrow \mathcal{\hat{T}}(2)|1\rangle$ for performing parallel distributed CNOT gates on spatially separated SC and SiV$^-$ qubits.


%


\end{document}